*Review*

# Passive Intermodulation at Contacts of Rough Conductors

Amir Dayan [1], Yi Huang [2] and Alex Schuchinsky [3,*]

[1] University of Liverpool, UK; Amir.Dayan@liverpool.ac.uk
[2] University of Liverpool, UK; Yi.Huang@liverpool.ac.uk
[3] University of Liverpool, UK; A.Schuchinsky@liverpool.ac.uk
* Correspondence: A.Schuchinsky@liverpool.ac.uk

**Abstract:** Passive intermodulation (PIM) is a niggling phenomenon that debilitates performance of modern communications and navigation systems. PIM products interfere with the information signals and cause their nonlinear distortion. The sources and basic mechanisms of PIM were studied in literature but PIM remains a serious problem of signal integrity. In this paper, the main sources and mechanisms of PIM generation by joints of good conductors are discussed. It is shown that the passive electrical, thermal and mechanical nonlinearities are intrinsically linked despite their distinctively different time scales. The roughness of the contact surfaces plays an important role in PIM generation by conductor joints. A review of the PIM phenomenology at contacts of the good conductors suggests that novel multiphysics models are necessary for the analysis and reliable prediction of PIM products generated by several concurrent nonlinearities of diverse physical nature.

## 1. Introduction

Continuously growing volume and speed of data transmission pose major challenges to existing and future wireless and satellite communications and navigation systems [1-7]. The stringent requirements for the integrity of information signals push the limits of the radio frequency (RF) hardware. Weak nonlinearities of passive devices such as antennas, filters, couplers and multiplexers at the RF front-end of the smart multi-radio base stations and user terminals generate spurious emission, corrupt information signals and debilitate the system performance [8-10]. Therefore, mitigation of nonlinear signal distortions is a major requirement to dynamically adjustable RF front-ends, their passive components and reconfigurable antennas [11-16].

Constituent materials and their contacts proved to be the main sources of passive nonlinearities in RF devices [17-19]. The state-of-the-art RF materials normally have low loss, high thermal conductivity and good mechanical properties. But when exposed to the high power of RF signals, they exhibit weakly nonlinear behaviour and generate frequency harmonics and passive intermodulation (PIM) products. PIM in passive "linear" devices is the result of mixing several high-power electromagnetic signals by weak nonlinearities of good conductors, their contacts and surface finish. The detrimental effect of PIM manifests itself in spurious emission, increased noise and distortion of the original information signals. PIM products are particularly harmful to radars, wireless and space communications systems and radio astronomy [5-7, 20, 21].

The PIM phenomenology has been studied for more than 40 years but still remains a nagging problem. The basic physical mechanisms of nonlinearities and PIM generation were explored in metal contacts [22-29], printed RF transmission lines [30-36], cable assemblies [17, 18, 31, 37-40], and antennas [21, 41-45]. The main sources of passive nonlin-earities include Metal-Insulator-Metal (MIM) junctions [22-24, 26-29], electro-thermal processes [37, 40, 46, 47], surface roughness [32, 48, 49] and contact mechanical deformations [49-53]. The advent of micro-electro-mechanical systems (MEMS) has sparked extensive investigations of the electrical contacts with rough surfaces and their RF performance. Whilst the mechanical properties of MEMS have been explored in great detail, see [50-53] and references therein, the existing models were developed for the linear devices operated with weak RF signals. The effect of the high RF power on the contacts and junctions of conductors with rough surfaces were studied only in the coaxial connectors [38-40] and waveguide flanges [54]. But the practical means of mitigating the nonlinear distortions and PIM in passive RF circuits remain predominantly semi-empirical [10, 55].

The analysis and prediction of PIM effects with the aid of the multiphysics models proved to be a challenging task. The published PIM studies are normally limited to a single dominant mechanism of nonlinearity. However, the stochastic nature of multifarious sources of weak nonlinearities and their intrinsic links dictates the need for multiphysics models and novel simulation tools. The diverse physical nature of PIM sources also requires that several concurrent mechanisms of nonlinearity[1] to be considered together.

PIM products generated by junctions of conductors with rough surfaces are usually characterised with the aid of the equivalent electrical circuit models. The lumped element model, proposed in [54], has been applied to the analysis of PIM in waveguide flanges where the physical dimensions of surface asperities are much smaller than the wavelengths of high-power RF signals. The parameters of this equivalent circuit depend not only on the electrical properties of the contact materials but also on their thermal and mechanical properties and the surface finish. Therefore, the concurrency of multiphysics effects is an essential feature of PIM in passive RF circuits and devices.

The PIM in coaxial connectors has recently been examined in [56-58] with the help of an equivalent circuit model from [54]. The relationship between an applied force and a contact linear resistance is evaluated in [59] using Meyer *empirical* law. It suggests that the contact resistance decreases as the square root of applied force when nonlinearity is approximated by a Taylor series. The simulation and measurement results presented in [59] are in fair agreement owing to the empirical description of the linear elements and data fitting for the nonlinear sources.

The main types of contact nonlinearities in the conductor joints and connectors can be cast in the three broad groups: (i) electrical, (ii) thermal, and (iii) mechanical. Their sources and interlinks are summarised in Figure 1 and include

- Charge tunnelling and diffusion at MIM junctions of conductor asperities.
- Current constriction at asperities of rough contact surfaces.
- Self-heating and thermal expansion of the contact surfaces and asperities.
- Electro-thermal effect due to ohmic losses in conductors and contacts.
- Asperity deformations of the conductors with rough surfaces and their contacts subjected to mechanical stresses, expansion and creep.

These nonlinearities exist in conductor joints concurrently and influence each other despite their notably different time scales. The underlying physical mechanisms are intrinsically linked as discussed in Sections 2-4 and are influenced by the layouts and profiles of the contact surfaces. The frequency and phasing of the high-power carrier signals also have a notable effect on the generated PIM products. For example, the positions of the hot spots in the antenna beamforming networks vary with the carriers' phasing. As the result, the observed pattern of PIM products is affected by a profile of the contact area, its deformations and heat flow. This implies that the locations of the PIM hot spots can change with the contact pressure that flattens sharp asperities and expands the contact area. This entails reduction of the current density at the MIM junctions and the lesser effect of their nonlinearity.

---

[1] The linearisation techniques used for power amplifiers are not suitable for mitigating PIM in the passive components of RF front-end and antennas due to absence of the real-time feedback required for signal predistortions.

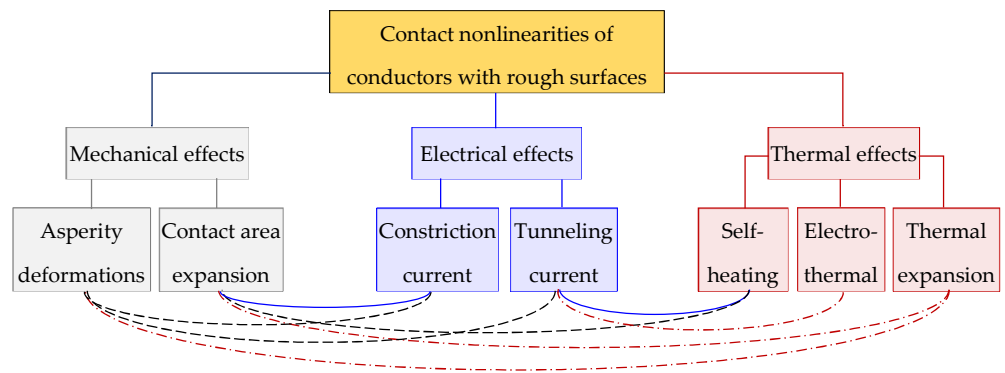

**Figure 1.** Main types of contact nonlinearities of rough conductor surfaces. Links between different mechanisms of nonlinearities are colour coded in relation to the primary source.

In this paper, we review the main types of nonlinearities at contact joints of good conductors and the related mechanisms of PIM generation. The effects of electrical nonlinearity in the conductor contacts are discussed in Section 2. The thermal and electro-thermal nonlinearities are the subject of Section 3. The effects of surface roughness and mechanical deformations of the conductor joints on PIM products are considered in Section 4. The main properties of different mechanisms of nonlinearities, their effect on PIM at contact joints of conductors with rough surfaces are summarized in Conclusion.

## 2. Electrical Nonlinearities of Conductor Joints

PIM at contacts of conductors with rough surfaces is a nonlinear multiphysics process with several distinct time scales, as illustrated by Figure 1. The fastest nonlinearities in contacts of good conductors are associated with the *electrical effects* of charge tunneling and current constriction at the MIM junctions [17]. The charges are funneled through the contact spots of rough surfaces as shown in Figure 2*a*. The current magnitude depends on the size of a contact spot and thickness of an insulting layer, which are determined by the applied pressure, local temperature and deformations of the surface asperities as illustrated by the links in Figure 1. The high-power carriers are mixed and modulated at the contact nonlinearities due to variations of the contact size, temperature and resistance [27, 38-40]. The thermal effects and mechanical deformations develop much slower than the oscillations of the RF carries and the tunneling current. But both fast and slow nonlinearities remain intrinsically coupled to each other despite their different time scales.

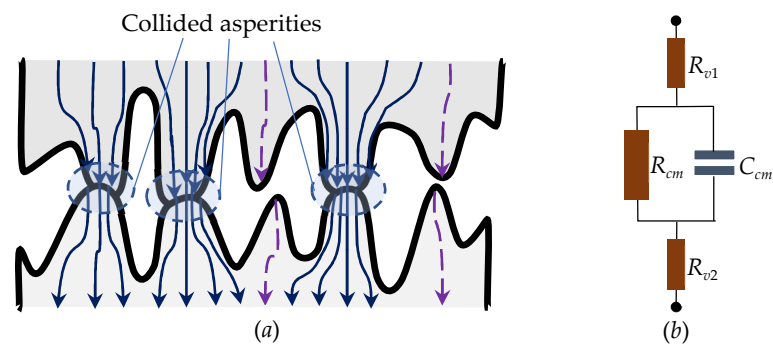

**Figure 2.** (*a*) Sketch of contacts between conductors with rough surfaces. Tips of the collided asperities are deformed and the constriction current is funneled through the contact areas. Only the displacement current (broken lines) is shown at the non-contact asperities. (*b*) Equivalent circuit of an asperity pair. $R_{v1}$ and $R_{v2}$ are the resistances of bulk conductors outside asperities. $R_{cm}$ and $C_{cm}$ are the resistance and capacitance of either the collided asperities or a pair of isolated asperities.

Current flow through the contact asperities of rough surfaces can be described by the equivalent circuit shown in Figure 2b, where $R_{v1}$ and $R_{v2}$ are the resistances of solid conductors outside of the asperities. The resistance and capacitance of a pair of compressed or non-touching asperities are represented by $R_{cm}$ and $C_{cm}$. The linear resistance $R_{cm}$ of the thin oxide film is very high even in contacts of good conductors. For example, $Al_2O_3$ film of thickness $\delta s$ = 1 nm, area 10 $\mu m^2$ and the macroscopic resistivity $\rho_{Al2O3}$ = 1·10$^{14}$ $\Omega$·cm has its Maxwell resistance $R_{cm}$ = 10$^{14}$ $\Omega$. Such a high value of $R_{cm}$ is typical for good insulators and suggests that the charge tunneling must be responsible for the resistance of conductor junctions with thin insulating films. It is also necessary to note that the capacitive reactance $j\omega C_{cm}$ of the conductor joint can noticeably influence the contact impedance. Indeed, the reactance of this junction is $1/j\omega C_{cm}$ = -$j$89.9 $\Omega$ that affects the impedance of the contact joint. Therefore, the resistance of the electrical contacts of rough conductor and their basic properties have to be examined in more detail.

*2.1. Contact and constriction resistances*

Contacts of conductor with rough surfaces contain the MIM and metal-to-metal junctions. Their resistance depends on the number of touching asperities, size of each contact spot and thickness of an insulating layer in MIM junctions [60]. The mechanical and thermal deformations of individual asperities determine an overall size of the contact area and its contact resistance. The resistance $R_c$ of a single contact spot with an equivalent radius $a$ is usually approximated as [61]

$$R_c = \frac{\rho_c}{2a}\left[f\left(\frac{\lambda}{a}\right) + \frac{8\lambda}{3\pi a}\right] \quad (1)$$

where $\rho_c$ is an average electrical resistivity of a pair of contact asperities, $\lambda$ is an electron free path and $f(\lambda/a)$ is an interpolation function describing a contribution of Maxwell resistance $R_M = \rho_c/2a$. The approximation of $f(\lambda/a)$ was proposed in [61] and has a maximum error less than 1% at any $\lambda/a$

$$f(\lambda/a) = \frac{1 + 0.83 \cdot \lambda/a}{1 + 1.33 \cdot \lambda/a} \quad (2)$$

The values of $f(\lambda/a)$ vary in a relatively narrow range between $f(\lambda/a) \approx 0.624$ at $\lambda \gg a$ and $f(\lambda/a) \approx 1$ at $\lambda \ll a$. The second terms in (1) represents Sharvin resistance [62]: $R_S = \frac{4\rho_c \lambda}{3\pi a^2}$ that is associated with the collision-free motion of charges. It plays an important role when the size $a$ of a contact spot is smaller than $\lambda$. Then Sharvin resistance $R_S$ can exceed Maxwell resistance $R_M$ and its contribution to the constriction current becomes significant.

Current constriction by asperities is an inherent feature of the contacts of conductors with rough surfaces [50, 54, 63]. The constriction resistance is determined by the number of the compressed asperities, sizes of their contact spots, and thicknesses of the insulating layer. The constriction current decreases, when the size of the contact spot is larger than the mean free path of electrons, $a > \lambda$, as evident from (2). Then the charge transport is predominantly diffusive, and it is determined by Maxwell resistance $R_M$, defined in (1). In the high-quality conductor contacts, e.g., in MEMS, the constriction current is much smaller than the conduction current and it is usually combined with the conduction or tunneling currents [64].

The effective resistance $R_u$ of the whole junction surface containing $M$ asperities can be averaged and approximated as suggested in [50]

$$R_u = \left(\sum_{m=1}^{M} R_{cm}^{-1}\right)^{-1} \approx \frac{\rho_{av}}{2a_{eff}} \left[ f\left(\frac{\lambda}{a_{eff}}\right) + \frac{8\lambda}{3\pi a_{eff}} \right] \quad (3)$$

where $M$ is the number of compressed asperity pairs, and $R_{cm}$ is defined by (1) and describes the contact resistance of $m^{th}$ pair of asperities. It was suggested in [50] that $R_u$ could be evaluated with (1) where $a$ and $\rho$ are replaced by an effective radius $a_{eff}$ and average resistivity $\rho_{av}$. Such an approximation given in (3) proved to be fairly accurate when the contact area is large at $a_{eff} > \lambda$. Then the contribution of Sharvin resistance is small and deviations of resistivities $\rho_{cm}$ from their average value $\rho_{av}$ remain small.

The contact resistances $R_{cm}$ of individual contact pairs and the whole ensemble, $R_u$, are the main electrical parameters used for characterising the performance of MEMS switches. They include the effects of both the electrical properties of the contact surfaces and the mechanical deformations. The electro-mechanical models of the contacts in the MEMS switches have been discussed in details in [50-52, 64].

## 2.2. Nonlinearity of MIM contacts

The tunnelling current in MIM contacts subjected to RF power is the main source of the fast nonlinearity [22-24]. When a thin insulating film of a few nanometre thick separates asperities of rough conductors, the tunnelling current flows through. The timescale of the tunnelling effects in the conductor junctions allows an efficient interaction of the high-power RF carriers with free charges. But when the thickness of the insulating layer exceeds the free path of electrons, charges cannot penetrate the potential barrier and the tunnelling current rapidly decays. The tunnelling current becomes practically negligible at the insulator between asperities thicker than 10 nm.

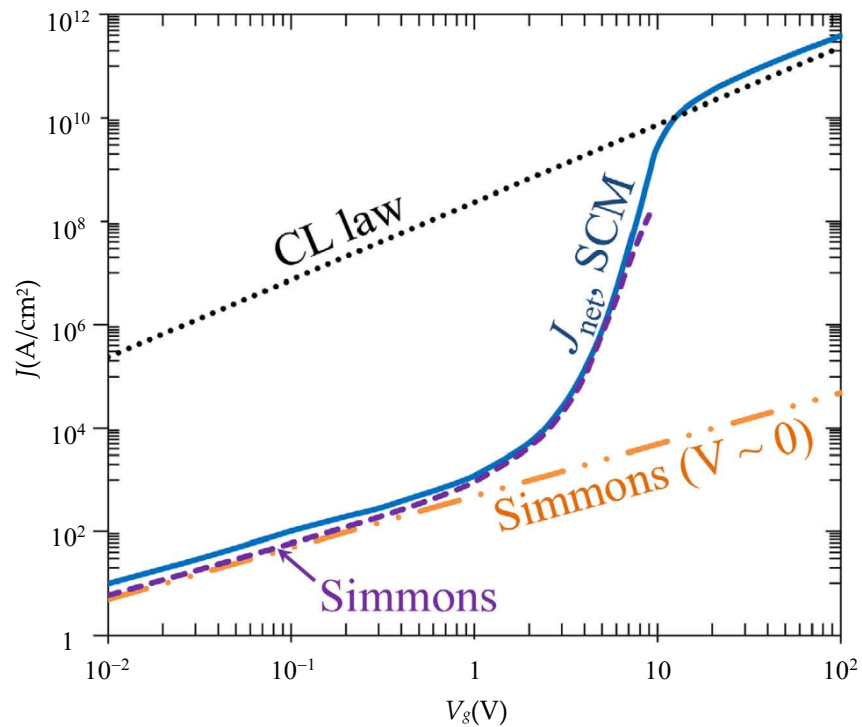

**Figure 3** [26]. Current density as a function of applied gap voltage $V_g$, for two gold (Au) electrodes (W = 5.1 eV) separated by a vacuum gap ($\varepsilon_r$ = 1) of width $D$ = 1 nm, at $T$ = 300 K. CL curve is for the classical Child-Langmuir law. $J_{net}$, SCM curve is calculated with the self-consistent model (SCM) described in [26]. Curves "Simmons" and "Simmons ($V \sim 0$)" are calculated with Simmons formulas [23] for general $V_g$ and $V_g \sim 0$, respectively.

The charge tunnelling between two *smooth* conductor surfaces was studied first by Somerfield and Bethe in the cases of the electric bias being either very weak or very strong. Holm has extended their model to the range of intermediate voltages [22] to characterise the tunnelling effect at a moderate electric field applied to MIM junction. For this purpose, Holm introduced an additional corrective term into the transition function, which determined the likelihood that electrons can pass through a thin insulator layer. However, Simmons showed that Holm's approximation was valid for the symmetric structure only and failed in asymmetric case. Simmons has analysed a general case of a potential barrier, assuming smooth variations of the barrier height and fairly small changes of its average value [23]. In the widely accepted Simmons model, the current density $J(V_g)$ in MIM junction is represented as

$$J(V_g) = J_0 \left[ \overline{\varphi} e^{-A\sqrt{\overline{\varphi}}} - (\overline{\varphi} + eV_g) e^{-A\sqrt{\overline{\varphi} + eV_g}} \right] \quad (4)$$

where $J_0 = \dfrac{e}{2\pi h (\beta \Delta s)^2}$ and $A = \dfrac{4\pi \beta \Delta s}{h}\sqrt{2m}$; $\overline{\varphi}$ is an average height of potential barrier inside an insulator layer, $V_g$ is the voltage between the contact conductors, $e$ is the electron charge, $h$ is the Plank constant, $\beta$ is the correction factor used in the approximation, $m$ is the electron mass, and $\Delta s$ is an effective thickness of insulator film which is usually smaller than the actual thickness. The value of the parameter $\beta$ is close to 1, and at insulator thicknesses $\Delta s \sim 4 - 5$ nm, the error of the approximation $\beta = 1$ is less than a few percent [23].

Accuracy of Simmons model (4) of MIM junction was recently examined for thin insulator films in [26-29]. The simulated characteristics of $J(V_g)$ in the MIM junction are shown in Figure 3 for the insulating film of thickness 1 nm. Comparison of Simmons model [23] with the self-consistent model (SCM) [26] demonstrates their good correlation. Simmons model also proved to be fairly accurate when the insulator thicknesses vary between 1 nm and 10 nm. It was also found that the equivalent resistivity $V_g/J$ inferred from (4) is slightly overestimated at sub-nanoscale $\Delta s$ and underestimated at $\Delta s > 5$-7 nm. Uncertainty of Simmons model in these cases increases to a few percent.

The effect of current crowding at the conductor edges and the nonuniform current distribution at the metal plates of partially overlapped parallel MIM contacts were modelled in [27-29]. The nonlinear tunnelling resistivity of MIM junction, defined as the ratio of voltage to current density, was evaluated in dependence of several parameters such as contact length, permittivity and thickness of an insulating layer, and the sheet resistance. It was observed that at the sub-nanometre insulator thicknesses 0.5 nm and 0.65 nm, the contact resistivity reduced and remained nearly linear at low applied voltages ($v < 0.3$ V). At voltages in the range $\sim 0.4$ V – 5 V, the slope of resistivity-voltage curves rapidly changes similar to Figure 3 and exhibits the essentially *nonlinear* behaviour of MIM junction. Thus, the MIM nonlinearity evidently represents a source of PIM that has not been considered in the literature, yet. The effects of conductor surface roughness, thickness and conductivity of the insulating layer on the MIM junction resistance are still debated in the literature.

*2.3. The Effect of Surface Roughness*

Surface roughness significantly affects the performance of electrical contacts in high frequency applications [49-52, 54]. Asperity random heights and patterns of the contact surfaces are never the same, even when the materials and fabrication process used are identical. Therefore, the statistical models have been used for simulating the contacts of rough surfaces. These techniques were applied to the characterisation of losses in printed circuit boards [48]. Surface roughness also proved to be an important factor influencing the performance of MEMS devices, and the contact phenomena in MEMS with rough conductors have been extensively studied in [50-53, 60, 63, 64].

Gaussian distribution of contact asperity heights with the standard deviation up to 20% from an average value was examined in [65], taking into account the effect of an insulator thickness on the contact resistance of rough surfaces. Using Brinkman-Dynes-Rowell model [66], it has been shown that a single thickness model reasonably estimates the surface conductance when the standard deviation of asperity heights remains within 5% of the average value, see Figure 4. However, as roughness increases, an average thickness of the insulator layer becomes smaller than its actual value used earlier for approximating the surface conductance. Also, at a larger standard deviation of thickness, a mean thickness decreases that makes a single-thickness approximation less accurate. Roughness also "flattens" the conductance as if the barrier were thinner. This effect is illustrated by the conductance curves for ideal single-thickness barriers (i.e., $\sigma \equiv 0$) presented for comparison (solid lines) in Figure 4 [65].

To relate a mean thickness to its actual value, inferred from the conductance measurements, the rules of thumb were suggested in [55]. However, accuracy and uniqueness of such fitting the normalized conductance values in a range of asperity heights remain a major problem as it was earlier pointed out in [67].

*2.4. Thermionic emission*

Thermionic emission is the process of electron discharge from the free metal surface. In good conductors, it is observed at very high temperatures, above 1000°C [68], when hot electrons can gain enough energy to break their bonding and escape from the conductor surface.

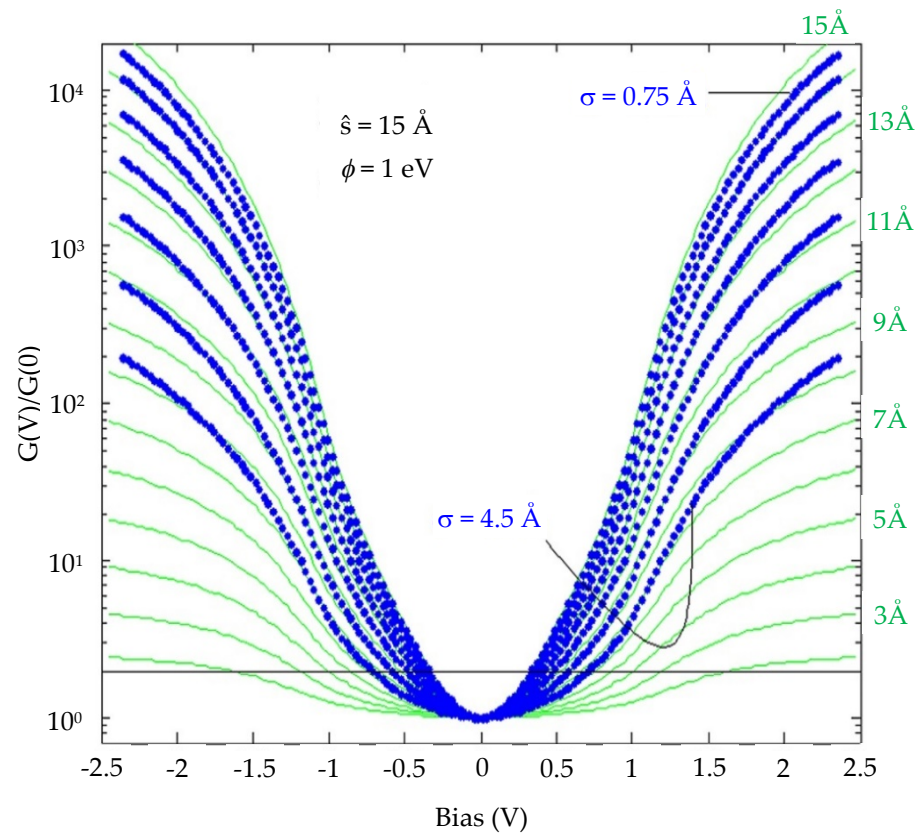

**Figure 4**. Bias dependence of normalized tunneling conductance, calculated for the barrier mean thickness $\hat{s}$ = 15Å, barrier height $\phi$ = 1.0 eV, and roughness $\sigma$ varying in the range of 5%–20% of $\hat{s}$ in steps of 2.5% (blue points), and single-thickness ($\sigma$≡0) conductance for 2–15Å in 1Å steps (solid green lines). Reproduced from [65] with the permission of AIP Publishing.

At operational temperatures of electronic devices, the thermionic emission current in joints of good conductors is negligible in comparison with the tunnelling current. It also decreases exponentially with the thickness of the insulating layer and becomes practically undetectable as the insulator layer is thicker than 5 nm. Thus, the thermionic emission current can be neglected in contacts of good conductors when the tunnelling current in MIM junctions exhibits the nonlinear behaviour as discussed in Section 2.2.

**3. Thermal Nonlinearities at Contacts of Good Conductors**

Thermal effects at conductor joints are associated with the main processes listed in Figure 1: (i) self-heating due to RF power dissipation in imperfect conductors, (ii) electro-thermal coupling in contact joints, and (iii) thermal expansion of the contact areas. They are linked intrinsically and influence other types of nonlinearities despite significant differences in the time scales of the underlying physical processes. Ambient temperature also has a notable impact on the effect of thermal nonlinearities [69]. The main properties and distinctive features of the thermal effects at contacts of conductors with rough surfaces are discussed below.

*3.1. Self-heating effect and nonlinearity of contact resistivity*

Heat generation is an inherent feature of electromagnetic (EM) wave interactions with conductors and their contact joints. This multiphysics process couples the electric and thermal domains as illustrated by Figure 5. Namely, the dissipative losses of high-power RF signals generate heat which, in turn, changes the resistance of conductors and causes their thermal expansion and mechanical deformations. These nonlinear processes are interlinked but the time scales of the thermal and mechanical processes notably differ.

High-power RF signals are attenuated by imperfect conductors due to their resistive losses. The dissipated power causes the conductor self-heating, and the changes of local temperature at point **x** alter a conductor resistivity $\rho_c(\mathbf{x})$. In a broad range of temperatures, $\rho_c(\mathbf{x})$ is well approximated by the linear temperature dependence

$$\rho_c(\mathbf{x}) = \rho_0 \left[ 1 + \alpha \cdot \delta T(\mathbf{x}) \right] \quad (5)$$

where $\rho_0$ is an average resistivity of the conductors at ambient temperature, $\alpha$ is the temperature coefficient of resistivity and $\delta T(\mathbf{x})$ is the local temperature increment due to RF heating of the contact

$$\delta T(\mathbf{x}) = \sum_n \delta q(\omega_n, \mathbf{x}) \cdot R_{\text{th,eq}}(\omega_n) \quad (6)$$

In (6), $\delta q(\omega_n, \mathbf{x})$ is the heat generated by the high-power carrier of frequency $\omega_n$ with current density $J_c(\omega_n, \mathbf{x})$ in an imperfect conductor [49]. Then

$$\delta q(\omega_n, \mathbf{x}) = \tfrac{1}{2} \text{Re}\left\{ Z_c(\omega_n, \mathbf{x}) \right\} \left| J_c(\omega_n, \mathbf{x}) \right|^2 \delta S \quad (7)$$

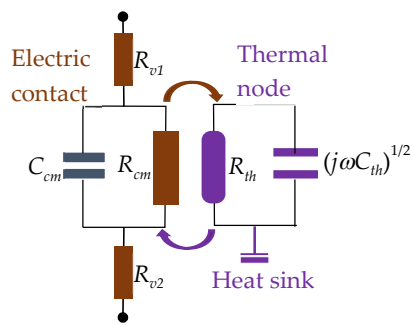

Figure 5. Compact model of a pair of the compressed asperities where the contact resistance $R_{cm}$ is coupled to a thermal node with the thermal resistance $R_{th}$ and thermal capacitance $C_{th}$.

where $Z_c(\omega_n, \mathbf{x})$ is a local impedance at carrier frequency $\omega_n$, and $\delta S$ is a differential area of the contact spot. An equivalent thermal resistance $R_{th,eq}(\omega_n)$ can be defined as [37]

$$R_{th,eq}(\omega_n) = \text{Re}\left\{\frac{R_{th}}{1 + R_{th}\sqrt{j\omega_n C_{th}}}\right\} \quad (8)$$

where $R_{th}$ and $C_{th}$ are the thermal resistance and thermal capacitance, respectively. They form a thermal node shown in Figure 5.

Simultaneous solution of (5)-(7) gives an approximation of the local resistivity

$$\rho_c(\mathbf{x}) = \rho_0\left[1 + \alpha\Delta_T(\mathbf{x}) + O(\alpha^2)\right] \quad (9)$$

where $\Delta_T(\mathbf{x}) = \delta S \sum_n \left[R_{th,eq}(\omega_n)\delta q(\omega_n, \mathbf{x})\right] = \tfrac{1}{2}\delta S \sum_n \left[R_{th,eq}(\omega_n)\text{Re}\{Z_{c0}(\omega_n)\}|J_c(\omega_n, \mathbf{x})|^2\right]$ (10)

is the resistivity increment proportional to the heat generated by the contact resistances Re{$Z_{c0}(\omega_n)$} at *ambient temperature*. Thus, (9), (10) show that the contact resistivity $\rho_c(\mathbf{x})$ depends on the squared current density *magnitudes* $|J_c(\omega_n,\mathbf{x})^2|$ of individual carriers and has the nonlinearity of Kerr type. An important feature of (9) is that $\rho_c(\mathbf{x})$ represents a local nonlinear resistivity, which may vary across the contact area due to inhomogeneity of the local thickness of the insulator in the MIM junction.

*3.2. Electro-thermal effect in contact joints and PIM generation*

Electro-thermal PIM (ET-PIM) is caused by self-heating due to conductor and dielectric losses [37]. ET-PIM has a distinct signature of the nonlinear coupling of electrical and thermal domains as illustrated by Figure 5. The basic mechanism of this nonlinearity is the thermal modulation of resistivity. In essence, the heat due to RF losses alters the resistance, which is heated by the high-power carriers, and this results in generation of PIM products. The effects of ET-PIM have been studied in the termination resistors [37, 47], printed TLs [36, 70] and thin-film coplanar waveguides with spatially inhomogeneous current distributions [35]. The developed theory of the ET-PIM and supporting experiments [36, 37, 70] have revealed that the baseband resistivity of conductors is modulated by the heat oscillations. The effect of resistivity variation on the skin depth due to modulation of the RF carriers was analyzed in [42, 43]. It enables both the development of the qualitative analytical model, which sheds the light on the principal mechanisms of ET-PIM generation, and an accurate assessment of signal distortion by ET-PIM in the full-wave EM simulations [33-36, 70].

Evaluation of the dissipative losses and the rate of self-heating are the critical steps in the ET-PIM analysis. While the skin effect is routinely modelled in the EM simulators when calculating losses of imperfect conductors, surface roughness is often ignored despite its proven major impact on the performance of printed circuit boards [48, 49] and MEMS switches [50-52]. The conductors with rougher surfaces exhibit not only higher RF losses and worth thermal performance but also generate a higher level of PIM [32].

The ET-PIM phenomenon was described first in [37] where a thermal modulation of the load resistance by two high-power RF signals was examined. A load with the size much smaller than the carrier wavelength has the resistance proportional to its temperature dependent resistivity $\rho_L(\delta T)$ where $\delta T$ is the temperature increment. Since $\rho_L(\delta T)$ has the same temperature dependence as $\rho_c(\mathbf{x})$ in (5), its modulation by the high-power carriers is described by (9). This basic model, shown in Figure 5, directly relates an instantaneous power of the electrical current in the load with resistance $R_{cm}$ to the heat flow at a thermal node with the thermal resistance $R_{th}$ and thermal capacitance $C_{th}$. Such a thermal system acts as a low pass filter as the thermal response of the resistive load is much slower than variations of the high-power RF signals of frequencies $\omega_1$ and $\omega_2$. However, when the frequency of thermal oscillations is comparable with an intermediate frequency $\Delta\omega = |\omega_2 - \omega_1|$ generated by the electro-thermal nonlinearity, the resulting temperature oscillations modulate the load resistance. Then the PIM products fall in the operational frequency band [35-37, 42, 46, 47] as illustrated by Figure 6. When the difference of the carrier frequencies exceeds several MHz, the thermal oscillations become too slow to follow the frequency $\Delta\omega$ and the magnitude of the ET-PIM products decreases [36, 37, 70].

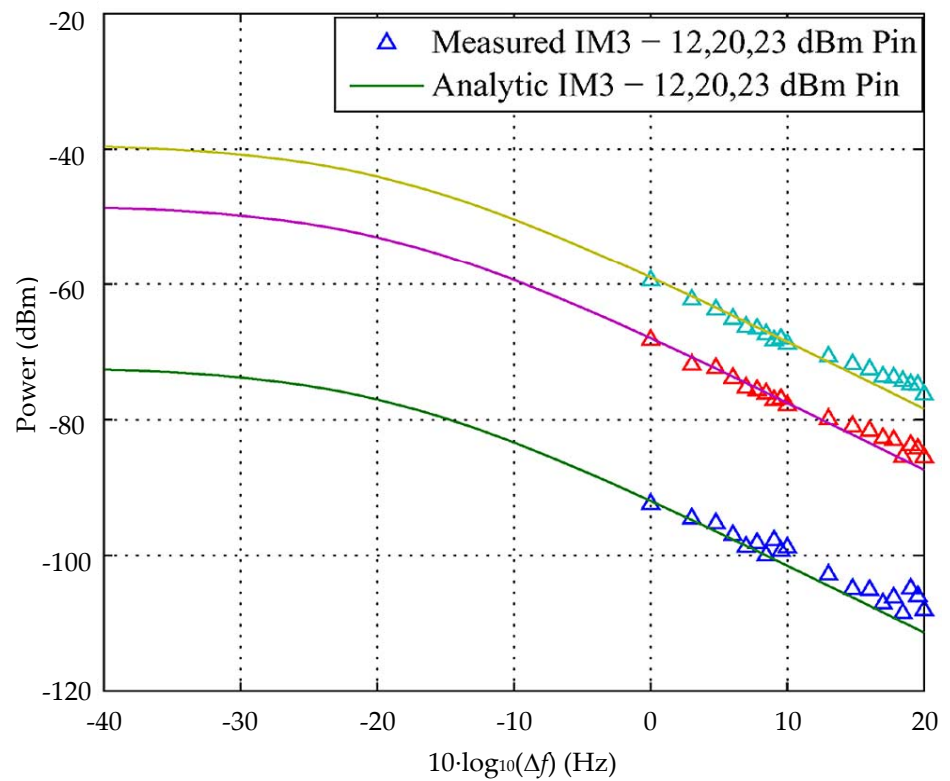

**Figure 6**. Analytic model prediction of 3rd order PIM products (solid lines) as compared with the measured IM3 products (triangles) for platinum at input power 12, 20, and 23 dBm versus carrier tone separation $\Delta f = \Delta\omega/2\pi$. © 2008 IEEE. Reprinted with permission from [37].

The distinctive feature of the ET-PIM is that the power slope of the PIM products versus a difference of carrier frequencies is only 10 dB per frequency decade [37]. Such a rate of the ET-PIM decay with the carrier frequency offset implies that ET-PIM products may have significantly stronger impact on the signal distortion in the broadband communications systems with a dense spectrum of the closely spaced high-power carriers.

The ET-PIM in contact joints is strongly affected by the surface roughness as the asperity height and size of a contact spot determine its resistance and reactance. The current constriction and contact heating at rough surfaces increase the spot temperature and asperity resistivity. The ET-PIM generated by joints of rough surfaces is related to their effective contact resistance $R_u$ defined in (3) and depends on the effective size $a_{eff}$ of a contact area and an averaged contact resistivity $\rho_{av}$ that increases with temperature as defined by (5). This results in the higher ET-PIM level at rougher surfaces that was observed in the printed circuits and conductor joints [32, 49, 64].

### 3.3. Thermal expansion of the contact area

The heat, generated by the high RF power, is dissipated in conductors and their joints and causes the thermal expansion and deformation of contact surfaces. As the result, the contact areas and the conductor resistivity $\rho_c(\mathbf{x})$ increase with temperature. The contact resistance $R_{cm}$ of asperities also grows with temperature, whilst the pace of its growth depends on the relation between the temperature coefficient of resistivity $\alpha$, the rate of the thermal expansion of the contact areas and asperity deformations as illustrated by Figure 1. An increase of the contact area is linked to the rate of heat flow from the contact spots but it progresses much slower. Therefore, the pace of establishing the thermal equilibrium is dictated by the lower speed of the asperity deformations.

The thermal effects at joints of conductors with rough surfaces are amplified by the current constriction by the contact spots. For example, as the temperature of an asperity increases due to current constriction, the thermal expansion of the compressed asperities somewhat reduces the constriction current density. But the growing contact pressure normally exceeds the yield stress of the heated asperities. As the result, they become susceptible to creep when subjected to compressive strain [71]. The effect of the contact pressure and the asperity deformations are dictated by the mechanical properties of the contact surfaces which are discussed next.

## 4. Mechanical Nonlinearities at Conductor Contacts

Mechanical deformations of electric contacts and conductor joints in connectors, MEMS switches, micro and nano electronic circuits have been extensively studied in the literature, see e.g., [49-53, 60, 64, 71] and references therein. The performance and longevity of contacts in RF devices proved to be affected by conductor surface roughness and stiffness. The high-frequency applications of MEMS have inspired the detailed studies of the effects of micro deformations on the contact resistivity, losses and reliability of conductor joints. The first models taking into account roughness of the conductor surfaces in microcontacts had been developed in [60, 64] and later refined in [50-53].

The contact resistance $R_c$ of asperities, defined in (1), is proportional to the electrical resistivity $\rho_c$ and depends on size $a$ of a contact spot. In a broad range of the contact pressure variation, $\rho_c$ of good conductors remains practically constant whereas the linear and nonlinear deformations of the contact spots become the dominant factors. It has also been observed in [64] that these deformations are weakly affected by thickness $t$ and permittivity $K$ of an insulating layer, as illustrated by Figure 7, if its energy level remains above the Fermi levels of the conductive surfaces. The asperity deformations and contact area expansion are also closely connected to the electrical and thermal effects discussed earlier and demonstrated by Figure 1.

### 4.1. Effect of asperity deformations on conductor resistivity

Contact resistances of rough surfaces calculated with the aid of (1) or (3) take into account the effects of an external pressure and deformations of the compressed asperities [50-54, 64, 71]. The factors affecting expansion of the contact areas and asperity compression include the mechanical stresses, elasticity and surface finish. At low pressure, the colliding asperities experience elastic deformations and their contact areas enlarge. When contact pressure increases, the compressed asperities experience plastic deformations and their contact regions are hardened. This results in slowing down the expansion of the contact spots and the change of the contact resistivity $\rho_c(\mathbf{x})$ [50, 60, 64]. Thus, the combined effect of the contact area enlargement and asperity hardening is governed by the relations between the contact pressure and contact resistance, cf. Figure 1. The nonlinear effects of strain hardening and softening on the resistivity $\rho(x)$ of a contact spot is accounted in [72]

$$\rho(\mathbf{x}) = \rho_c(\mathbf{x})\left(1 + \frac{\varepsilon_p}{\varepsilon_{ref}}\right)^q \left[1 - \exp\left(-\frac{Q}{kT}\right)\right] \tag{11}$$

where $\rho_c(\mathbf{x})$ is the temperature dependent resistivity of a contact, defined in (5), $\varepsilon_p$ is plastic strain, $\varepsilon_{ref}$ is a reference strain, $q$ is a material dependent parameter, $Q$ is the activation energy for the mechanism of relaxing the stored dislocations, $T$ is the absolute temperature and Boltzmann constant $k = 1.38 \times 10^{-23}$ J/K.

In addition to altering the contact resistance, heat softens contact asperities [50-53] and increases their plastic and creep deformations. Very slow creep deformations have been examined in micro-contacts with a low current in [50], where the strain-rate $\dot{\varepsilon}$ is described by a power law dependence on the stress $\sigma$ as

$$\dot{\varepsilon} = A\sigma^p \exp\left(-\frac{Q_c}{kT}\right) \tag{12}$$

where $A$ is a parameter defined by the material properties and the creep mechanism, $\sigma$ is the stress, $Q_c$ is the activation energy for creep. The stress exponent $p$ used in (12) usually varies between 3 and 10, depending on the material composition.

When contuct joints are exposed to the high RF power, the contact area of asperities increases further due to its thermal expansion. But the conductor resistivity $\rho_c(x)$ also grows due to Joule heat as discussed in Section 3. These two effects counteract each other as the rise of contact temperature increases both $\rho_c(x)$ and the dissipative losses, whilst enlargement of the contact area reduces the resistance and a current density. Joule heat, generated by the current flowing through contact asperities, increases the temperature $T$ of the contact spots. The temperature dependence on the contact voltage $V_c$ was examined in [73, 74] for conductor joints with the characteristic size $a$ of the contact spot considerably larger than the electron mean free path $\lambda$. Since the boundary scattering has a minor effect on the self-heating of the contact spots in this case, the contact temperature is evaluated with the asperity heating model [50]

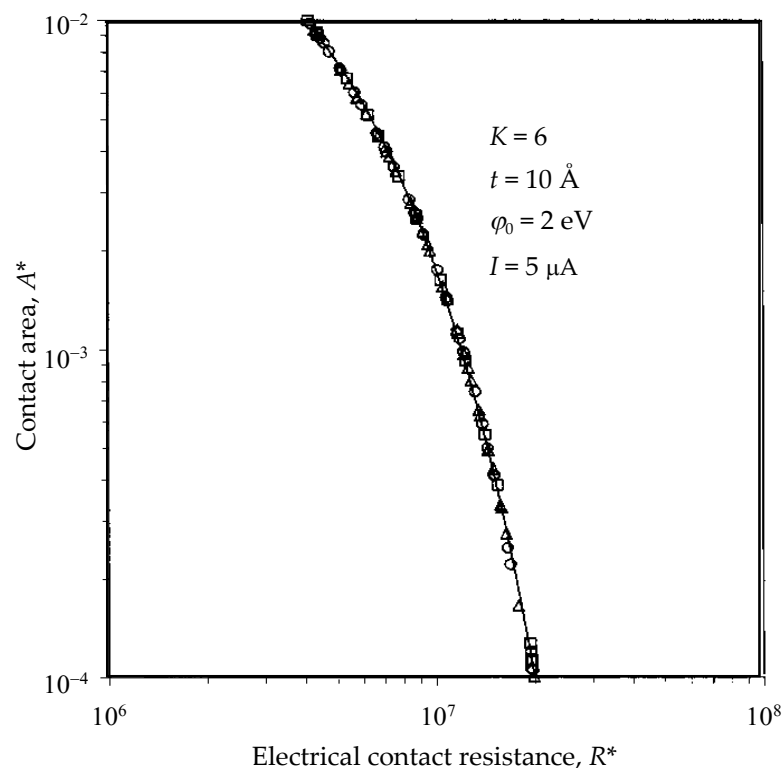

**Figure 7.** Dimensionless contact area $A^*$ vs dimensionless electrical contact resistance $R^*$ for contacting rough surfaces with different mechanical properties and fractal parameters ($K = 6$, $t = 10$ Å, $\varphi_0 = 2$ eV, and $I = 5$ mA). Reproduced from [64] with the permission of AIP Publishing.

$$T = \sqrt{T_0^2 + \frac{\gamma}{4L} \frac{V_c^2}{\xi(\lambda/a)}} \quad (13)$$

where $T_0$ is ambient temperature, $\gamma$ is the scaling parameter of the asperity distribution, $L = 2.45 \cdot 10^{-8}$ W·Ω/K$^2$ is Lorentz number, $V_c$ is the contact voltage, $\xi(x) = 1 + \frac{8}{3\pi} \frac{x}{f(x)}$ and $f(x)$ is defined in (2). It is interesting to note in (13) that $T$ depends on $\rho(\mathbf{x})$ only implicitly, through contact voltage $V_c$.

The slow deformations of contact asperities in conductor joints also affect the temperature dependent resistivity $\rho(\mathbf{x})$ which is related to the material stress and strain. Variations of $\rho(\mathbf{x})$ due to the contact deformations are much slower than the charge tunnelling and thermal effects. Therefore, the mechanical effects in the RF contacts can be considered in the quasi-static approximation that significantly simplifies the multiphysics analysis of the complex conductor joints.

*4.2. Contact area expansion*

Pressure, applied to the contacts of conductors with rough surfaces, causes asperity deformations. The compressed asperities expand laterally, and their contact areas increase with the applied pressure and mechanical stress, depending on the material stiffness and surface coatings. When exposed to the RF power, the contact spots are heated due to conductor losses and the compressed asperities spread gradually to reduce their strain. The thermal expansion and softening of the contact spots also cause plastic deformations and creep of the colliding asperities. These effects were examined in [71] for RF MEMS with frustoconical contact asperities. At a small radius $r_1$ of the undeformed frustum tip, contact radius $r_c(t)$ of the compressed asperity varies with time $t$ and is approximated as [71]

$$r_c(t) = \left[ r_1^{1/\alpha} + t \frac{LC}{\alpha \tan\beta} \left(\frac{F}{\pi}\right)^{(1/\alpha - 1)/2} \exp\left(-\frac{Q_c}{kT}\right) \right]^\alpha \quad (14)$$

where $L$ is asperity initial height, $C$ is a constant dependent on the contact material and creep mechanism, $\beta$ is a slant angle of the asperity frustum, $F$ is contact load, $Q_c$ is the activation energy for creep, $\alpha = 1/(1+2p)$, and $p$ is the stress exponent defined in (12). The value of $\alpha$ is specifically related to the creep coefficient, and it is fitted to the experimental data. Values of $\alpha$ are usually small, and $\alpha < 0.1$ in the example presented in [71].

Maxwell contact resistance $R_M$ of conical frustums with the time-dependent radius $r_c(t)$ and resistivity $\rho_c$ of the contact spot can be calculated with (1). The simulated and measured time-dependent resistances $R_M(t)$ are shown in Figure 8. Approximated by (1) with constant $\rho_c$ and the time variable contact size $r_c(t)$ defined in (14), $R_M(t)$ fits the measurement results very well when the switch is closed the first time. When the contacts are closed second and third times (middle and top plots), the measured resistances $R_M(t)$ are flatter than the calculated ones at the first $\delta t = 4$ s and $\delta t = 9$ s, respectively. But $R_M(t)$ are well correlated with the simulations later on. It is also necessary to note that Sharvin resistance showed no visible effect on the contact resistance and its time variations here. This is a manifestation of the incommensurate time scales of the effects of the charge tunneling and the mechanical deformation in the contacts of conductor with rough surfaces.

The dynamics of MEMS contacts was alternatively examined in [51, 52] where the contact area was evaluated in the frequency domain. This approach required less computational resources but its accuracy was lower, and it did not allow the analysis of the temporal variations of the contact area and its resistance. These works were applicable to the very slow contact deformations that allowed calculations of the size of the expanded contact area, which determined the contact resistance only.

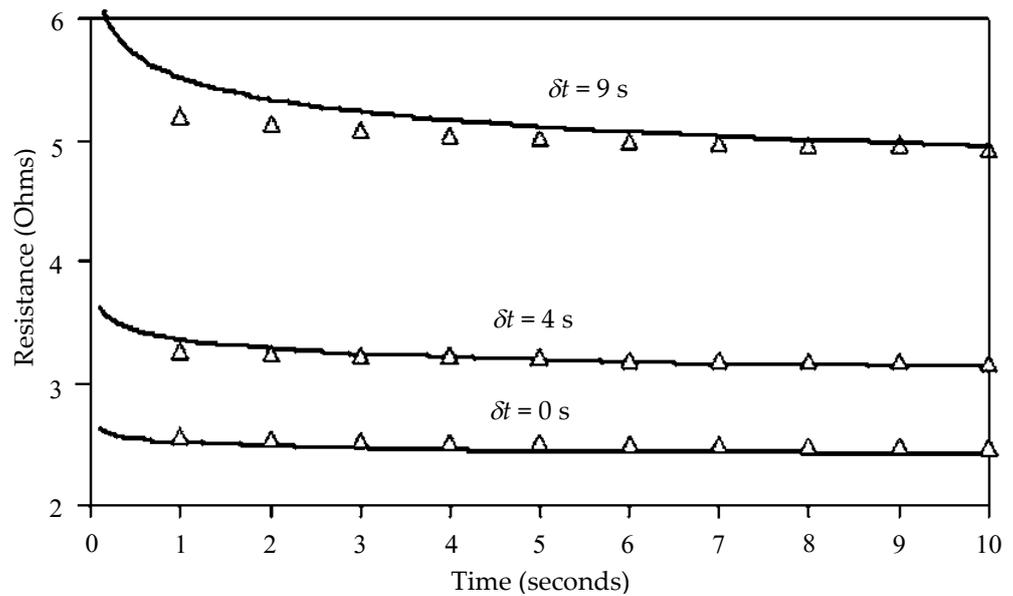

**Figure 8**. Symbols are the experimental data; solid lines are the fit of equation $\rho = At^{-a} + B$ to each entire data set used in [71]. Reproduced from [71] with the permission of AIP Publishing.

## 5. Conclusions

The main mechanisms of dissipative losses and sources of passive nonlinearities in contacts and joints of conductors with rough surfaces are reviewed. The discussed physical mechanisms are cast in the three main groups of (i) electrical, (ii) thermal, and (iii) mechanical effects. It is shown that all these mechanisms are intrinsically linked but their time scales differ significantly. Namely, the electromagnetic interactions at MIM junctions are very fast and can follow the pace of the RF signals. The thermal processes are much slower being limited by the speed of heat flow in the conductor contacts. And the mechanical deformations develop even slower. It is emphasised that roughness of the contact surfaces considerably affects losses and nonlinearity in contact joints, especially at RF frequencies. The main sources of the passive nonlinearities at the contacts of rough surfaces have been discussed in the context of their effect on the RF performance of the joints of good conductors.

An important outcome of this study is a demonstration that types of the electrical, thermal and mechanical contact nonlinearities are very different in spite of being linked intrinsically. Namely, the MIM nonlinearity is of the exponential type, whereas the electro-thermal nonlinearity is of Kerr type. The nonlinearities of mechanical deformations are of the mixed type described by the combination of the exponential and algebraic dependencies. This implies that the analysis of PIM in contact joints of good conductors requires the multiphysics models taking into account multiple sources of nonlinearities.

**References**

1. Kanhere, O.; Rappaport, T. S. Position Location for Futuristic Cellular Communications: 5G and Beyond, *IEEE Commun. Mag.*, **2021**, *59*, pp. 70-75.
2. Saad, W.; Bennis, M.; and Chen, M. A Vision of 6G Wireless Systems: Applications, Trends, Technologies, and Open Research Problems," *IEEE Network*, **2020**, *3*, pp. 134–142.

3. Giordani M. et al. Toward 6G Networks: Use Cases and Technologies, *IEEE Commun. Mag.*, **2020**, *2*, pp. 244-251.
4. Rappaport, T. S. et al. Wireless Communications and Applications Above 100 GHz: Opportunities and Challenges for 6G and Beyond, *IEEE Access*, **2019**, *7*, pp. 78,729–57.
5. Wei T.; Feng, W.; Chen, Y.; Wang, C. -X.; Ge, N.; Lu, J. Hybrid Satellite-Terrestrial Communication Networks for the Maritime Internet of Things: Key Technologies, Opportunities, and Challenges, *IEEE Internet of Things J*, **2021**, *11*, pp. 8910-8934.
6. Kodheli O. *et al.* Satellite Communications in the New Space Era: A Survey and Future Challenges, *IEEE Commun. Surveys & Tutorials*, **2021**, *1*, pp. 70-109.
7. Alkendi, Y.; Seneviratne L.; Zweiri, Y. State of the Art in Vision-Based Localization Techniques for Autonomous Navigation Systems, *IEEE Access*, **2021**, *9*, pp. 76847-76874.
8. Waheed M. Z. *et al*. Passive Intermodulation in Simultaneous Transmit–Receive Systems: Modeling and Digital Cancellation Methods, *IEEE Trans. Microw. Theory and Techn.*, **2020**, *9*, pp. 3633-3652.
9. Kozlov D. *et al*. Practical Mitigation of Passive Intermodulation in Microstrip Circuits, *IEEE Trans. on Electromagnetic Compatibility*, **2020**, *1*, pp. 163-172.
10. Zhang, S.; Zhao, X.; Cao, Z.; Zhang, K.; Gao, F.; He, Y. Experimental Study of Electrical Contact Nonlinearity and its Passive Intermodulation Effect," *IEEE Trans. Comp., Packag. Manufact. Techn.*, **2020**, *3*, pp. 424-434, doi: 10.1109/TCPMT.2019.2955283.
11. Gard, K. G.; Larson, L. E.; Steer, M. B. The impact of RF front-end characteristics on the spectral regrowth of communications signals," *IEEE Trans. Microw. Theory and Techn.*, **2005**, *6*, pp. 2179-2186, doi: 10.1109/TMTT.2005.848801.
12. Grimm, M.; Allén, M.; Marttila, J.; Valkama, M.; Thomä, R. Joint Mitigation of Nonlinear RF and Baseband Distortions in Wideband Direct-Conversion Receivers, *IEEE Trans. Microw. Theory and Techn.*, **2014**, *1*, pp. 166-182, doi: 10.1109/TMTT.2013.2292603.
13. Vansebrouck, R.; Jabbour, C.; Jamin, O.; Desgreys, P. Fully-Digital Blind Compensation of Non-Linear Distortions in Wideband Receivers, *IEEE Trans. Circuits and Systems I*, **2017**, *8*, pp. 2112-2123, doi: 10.1109/TCSI.2017.2694406.
14. Yang, P. et al., 6G Wireless Communications: Vision and Potential Techniques, *IEEE Network*, **2019**, *4*, pp. 70–75.
15. Zardi, F.; Nayeri, P.; Rocca, P.; Haupt, R. Artificial Intelligence for Adaptive and Reconfigurable Antenna Arrays: A Review, *IEEE Antennas and Propag. Mag*, **2021**, *3*, pp. 28-38, doi: 10.1109/MAP.2020.3036097.
16. Morales-Cespedes, M.; Vandendorpe L.; Armada, A. G. Interference Management for K-Tier Networks without CSIT based on Reconfigurable Antennas, *IEEE Trans. Communications*, **2021**, doi: 10.1109/TCOMM.2021.3116246.
17. Stauss, G. H. Intrinsic sources of IM generation, Naval Res. Lab., Washington, DC, USA, Tech. Rep. 4233, **1980**, ch. 5, pp. 65–82.
18. Foord, A. P.; Rawlins, A. D. A study of passive intermodulation interference in space RF hardware, Final Rep., **1992,** Univ. Kent Canterbury, Canterbury, U.K., ESTEC Contract 111036.
19. Seron, D.; Collado, C.; Mateu, J.; O'Callaghan, J. M. Analysis and simulation of distributed nonlinearities in ferroelectrics and superconductors for microwave applications. *IEEE Trans. Microw. Theory Techn.* **2006**, *3*, pp. 1154-1160, doi: 10.1109/TMTT.2005.864110.
20. Boyhan, J. W.; Henzing, H. F.; Koduru, C. Satellite passive intermodulation: systems considerations, *IEEE Trans. Aerospace and Electronic Systems*, **1996**, *3*, pp. 1058-1064, doi: 10.1109/7.532264.
21. Bolli, P.; Selleri, S.; Pelosi, G. Passive intermodulation on large reflector antennas, *IEEE Antennas and Propagation Mag.*, **2002**, *5*, pp. 13-20, doi: 10.1109/MAP.2002.1077773.
22. Holm, R. The electric tunnel effect across thin insulator films in contacts. *J. Applied Physics*, **1951**, 5, p. 569-574.
23. Simmons, J.G., Generalized formula for the electric tunnel effect between similar electrodes separated by a thin insulating film. *J. Applied Physics*, **1963**, *6*, p. 1793-1803.
24. Bond, C. D.; Guenzer, C. S.; Carosella, C. A. Intermodulation generation by electron tunneling through aluminum-oxide films, *Proc. IEEE*, **1979**, *12*, pp. 1643-1652, doi: 10.1109/PROC.1979.11544.
25. Yamamoto, Y.; Kuga, N. Short-Circuit Transmission Line Method for PIM Evaluation of Metallic Materials, *IEEE Trans. Electromagnetic Compatibility*, **2007**, *3*, pp. 682-688, doi: 10.1109/TEMC.2007.902404.
26. Zhang, P. Scaling for quantum tunneling current in nano- and subnano-scale plasmonic junctions. Sci. Rep., **2015**, *5*, 9826, doi: 10.1038/srep09826.
27. Banerjee, S.; Zhang, P. A generalized self-consistent model for quantum tunneling current in dissimilar metal-insulator-metal junction. *AIP Advances*, **2019**, *9*, 085302, doi: 10.1063/1.5116204.
28. Banerjee, S.; Luginsland, J.; Zhang, P. A two dimensional tunneling resistance transmission line model for nanoscale parallel electrical contacts. *Sci. Rep.*, **2019**, *9*, 14484, doi: 10.1038/s41598-019-50934-2.
29. Banerjee, S.; Luginsland, J.; Zhang, P. Interface Engineering of Electrical Contacts, *Phys. Rev. Applied*, **2021**, *15*, 064048, doi: 10.1103/PhysRevApplied.15.064048.
30. Auld, B. A.; Didomenico Jr., M.; Pantell, R. H. Traveling-wave harmonic generation along nonlinear transmission lines, *J. Appl. Phys.*, **1962**, 12, pp. 3537–3545, doi: 10.1063/1.1702443.
31. Bayrak, M.; Benson, F. Intermodulation products from nonlinearities in transmission lines and connectors at microwave frequencies. *Proc. IEE*, **1975**, *4*, pp. 361-367, doi: 10.1049/piee.1975.0101.
32. Schuchinsky, A. G.; Francey, J.; Fusco, V. F. Distributed sources of passive intermodulation on printed lines, Proc. IEEE Antennas Propag. Soc. Int. Symp., **2005**, 4B, pp. 447-450, doi: 10.1109/APS.2005.1552847.
33. Zelenchuk, D. E.; Shitvov, A. P.; Schuchinsky, A. G.; Fusco, V.F. Passive Intermodulation in Finite Lengths of Printed Microstrip Lines, *IEEE Trans. Microw. Theory Techn.* **2008**, *11*, pp. 2426-2434, doi: 10.1109/TMTT.2008.2005886.

34. Shitvov, A. P.; Zelenchuk, D. E.; Schuchinsky, A. G.; Fusco, V.F. Passive Intermodulation Generation on Printed Lines: Near-Field Probing and Observations", *IEEE Trans. Microw. Theory Techn*. **2008**, *12*, pp. 3121-3128, doi:10.1109/TMTT.2008.2007136.
35. Mateu, J. et al. Third-order intermodulation distortion and harmonic generation in mismatched weakly nonlinear transmission lines, *IEEE Trans. Microw. Theory Techn.*, **2009**, *1*, pp. 10–18, doi: 10.1109/TMTT.2008.2009083.
36. Wilkerson, J.R.; Lam, P. G.; Gard K. G.; Steer, M. B. Distributed passive intermodulation distortion on transmission lines. *IEEE Trans. Microw. Theory Techn*., **2011**, *5*, pp. 1190-1205, doi: 10.1109/TMTT.2011.2106138.
37. Wilkerson, J.R.; Gard K. G.; Schuchinsky, A. G.; Steer, M. B. Electro-thermal theory of intermodulation distortion in lossy microwave components, *IEEE Trans. Microw. Theory Techn*. **2008**, *12*, pp. 2717-2725, doi: 10.1109/TMTT.2008.2007084.
38. Jin, Q.; Gao, J.; Flowers, G. T.; Wu, Y.; Huang, H. Passive Intermodulation Models of Current Distortion in Electrical Contact Points," *IEEE Microw. Wireless Comp. Lett.*, **2019**, *3*, pp. 180-182, doi: 10.1109/LMWC.2019.2894799.
39. Jin, Q.; Gao, J.; Bi, L.; Zhou, Y. The Impact of Contact Pressure on Passive Intermodulation in Coaxial Connectors," *IEEE Microw. Wireless Comp. Lett.*, **2020**, *2*, pp. 177-180, doi: 10.1109/LMWC.2019.2957983.
40. Chen, X.; Wang, L.; Pommerenke, D.; Yu, M. Passive Intermodulation on Coaxial Connector Under Electro-Thermal-Mechanical Multiphysics, *IEEE Trans. Microw. Theory Techn*. **2021**, doi: 10.1109/TMTT.2021.3103981.
41. Hienonen, S.; Vainikainen, P.; Raisanen, A. V. Sensitivity measurements of a passive intermodulation near-field scanner, *IEEE Antennas Propag. Mag.*, **2003**, *4*, pp. 124-129, doi: 10.1109/MAP.2003.1241323.
42. Wilkerson, J.R.; Kilgore, I. M.; Gard K. G.; Steer, M. B. Passive Intermodulation Distortion in Antennas, *IEEE Trans. Antennas Propag.*, **2015**, *2*, pp. 474-482, doi: 10.1109/TAP.2014.2379947.
43. Wang, M.; Kilgore, I. M.; Steer, M. B.; Adams, J. J. Characterization of Intermodulation Distortion in Reconfigurable Liquid Metal Antennas, *IEEE Antennas Wireless Propag. Lett.*, **2018**, *2*, pp. 279-282, doi: 10.1109/LAWP.2017.2786078.
44. Wu, D.; Xie, Y.; Kuang, Y.; Niu, L. Prediction of Passive Intermodulation on Mesh Reflector Antenna Using Collaborative Simulation: Multiscale Equivalent Method and Nonlinear Model," *IEEE Trans. Antennas Propag.* **2018**, *3*, pp. 1516-1521, doi: 10.1109/TAP.2017.2786304.
45. Yong, S.; Yang, S.; Zhang, L.; Chen, X.; Pommerenke, D. J.; Khilkevich, V. Passive Intermodulation Source Localization Based on Emission Source Microscopy," *IEEE Trans. Electromagnetic Compat.* **2020**, *1*, pp. 266-271, doi: 10.1109/TEMC.2019.2938634.
46. Wilcox, J.; Molmud, P. Thermal Heating Contribution to Intermodulation Fields in Coaxial Waveguides, *IEEE Trans. Communications*, **1976**, *2*, pp. 238-243, doi: 10.1109/TCOM.1976.1093272.
47. Steer, M. B.; Williamson, T. G.; Wetherington, J.; Wilkerson, J.; Aaen, P.; Schuchinsky, A. G. Power and temperature dependence of passive intermodulation distortion, 22nd Int. Microwave and Radar Conf. (MIKON), **2018**, pp. 428-431, doi: 10.23919/MIKON.2018.8405245.
48. Guo, X.; Jackson, D.R.; Koledintseva, M.Y.; Hinaga, S.; Drewniak, J.L.; Ji Chen, An analysis of conductor surface roughness effects on signal propagation for stripline interconnects, *IEEE Trans. Electromagnetic Compatibility*, **2014**, *3*, pp. 707-714, doi: 10.1109/TEMC.2013.2294958.
49. Ansuinelli, P.; Schuchinsky, A. G.; Frezza F., Steer, M. B. Passive intermodulation due to conductor surface roughness, *IEEE IEEE Trans. Microw. Theory Techn*., **2018**, *2*, pp. 688-699, doi: 10.1109/TMTT.2017.2784817.
50. Rezvanian, O.; Zikry, M.A.; Brown, C.; Krim, J. Surface roughness, asperity contact and gold RF MEMS switch behavior. *J. of Micromechanics and Microengineering*. **2007**, *10*, pp. 2006-2015, doi: 10.1088/0960-1317/17/10/012.
51. Jackson, R.L.; Streator, J.L. A multi-scale model for contact between rough surfaces. *Wear*, **2006**, *11-12*, pp. 1337-1347, doi: 10.1016/j.wear.2006.03.015.
52. Jackson, R. L.; Crandall, E. R.; Bozack, M. J. Rough surface electrical contact resistance considering scale dependent properties and quantum effects, *J. Appl. Phys.*, **2015**, 117, pp. 195101, doi: 10.1063/1.4921110.
53. Zhai, C.; Hanaor, D.; Proust, G.; Brassart, L.; Gan Y. Stress-dependent electrical contact resistance at fractal rough surfaces, *J. Eng. Mech.*, **2017**, *3*, B4015001, doi: 10.1061/(ASCE)EM.1943-7889.0000967.
54. Vicente, C.; Hartnagel, H.L. Passive-intermodulation analysis between rough rectangular waveguide flanges. *Microw. Theory Techn.*, **2005**, *8*, p. 2515-2525, doi: 10.1109/TMTT.2005.852771.
55. Yang, H.; Zhu, L.; Gao, F.; Fan, J.; Wen, H. Measurement and analysis of passive intermodulation induced by additional impedance in loose contact coaxial connector, *IEEE Trans. Electromagnetic Compat.*, **2019**, *6*, pp. 1876-1883, doi: 10.1109/TEMC.2019.2892458.
56. Zhang, K.; Li, T.; Jiang, J. Passive intermodulation of contact nonlinearity on microwave connectors. *IEEE Trans. Electromagnetic Compatibility*, **2018**, 2, p. 513-519, doi: 10.1109/TEMC.2017.2725278.
57. Jin, Q.; Gao, J.; Flowers, G. T.; Wu, Y.; Xie, G. Modeling of passive intermodulation with electrical contacts in coaxial connectors, *IEEE Trans. Microw. Theory Techn*. **2018**, *9*, pp. 4007-4016, doi: 10.1109/TMTT.2018.2838147
58. Chen, X.; He, Y.; Yu, M.; Pommerenke, D. J.; Fan, J. Empirical Modeling of Contact Intermodulation Effect on Coaxial Connectors, *IEEE Trans. Instrumentation Measurement*, **2020**, *7*, pp. 5091-5099, doi: 10.1109/TIM.2019.2957869.
59. Chen, X., et al., Analytic passive intermodulation behavior on the coaxial connector using Monte Carlo approximation. *IEEE Trans. Electromagnetic Compatibility*, **2018**, *5*, p. 1207-1214, doi: 10.1109/TEMC.2018.2809449.
60. Kogut, L.; Komvopoulos, K. Analytical current-voltage relationships for electron tunneling across rough interfaces. *J. Appl. Phys.*, **2005**, 97, 073701, doi: 10.1063/1.1866472.
61. Nikolic, B.; Allen, P.B. Electron Transport through a circular constriction, *Phys. Rev. B*, **1999**, *6*, pp. 3963-3969, doi: 10.1103/PhysRevB.60.3963.

62. Sharvin, Y.V. On the possible method for studying Fermi surfaces. *Sov. Phys. JETP*, **1965**, 21, 3, pp. 655-656.
63. Jensen, B. D.; Chow, L. L.-W.; Huang, K.; Saitou, K.; Volakis, J. L.; Kurabayashi, K. Effect of nanoscale heating on electrical transport in RF MEMS switch contacts, *J. Microelectromechanical Systems*, **2005**, *5*, pp. 935-946, doi: 10.1109/JMEMS.2005.856653.
64. Kogut, L.; Komvopoulos, K. Electrical contact resistance theory for conductive rough surfaces separated by a thin insulating film." *J. Appl. Phys.*, **2004**, *2*, pp. 576–585, doi: 10.1063/1.1629392.
65. Miller, C. W.; Li, Zhi-Pan; Åkerman, J.; Schuller, I. K. Impact of interfacial roughness on tunneling conductance and extracted barrier parameters, *Appl. Phys. Lett.*, **2007**, 90, 043513, doi: 10.1063/1.2431443.
66. Brinkman, W. F.; Dynes, R. C.; Rowell, J. M. Tunneling Conductance of Asymmetrical Barriers, *J. Appl. Phys.*, **1970**, 41, pp. 1915-1921, doi: 10.1063/1.1659141
67. Crinon, E.; Evans, J.T. The effect of surface roughness, oxide film thickness and interfacial sliding on the electrical contact resistance of aluminium, *Materials Sci. Eng.: A*, **1998,** 1–2, pp 121-128. doi: 10.1016/S0921-5093(97)00508-X.
68. Go, D. B.; Haase, J. R.; George, J.; Mannhart, J.; Wanke, R.; Nojeh, A.; Nemanich, R. Thermionic energy conversion in the twenty-first century: Advances and Opportunities for Space and Terrestrial Applications, Frontiers in Mechanical Engineering, **2017**, 13 pp., doi: 10.3389/fmech.2017.00013.
69. Jin, Q.; Gao, J.; Flowers, G. T.; Wu, Y.; Xie, G.; Bi, L. Effects of environmental temperature on passive intermodulation in electrical connectors, *IEEE Trans. Components, Packaging and Manufact. Techn.*, **2020**, 12, pp. 2008-2017, doi: 10.1109/TCPMT.2020.3028432.
70. Rocas, E.; Collado, C.; Mateu, J.; Orloff, N.; O'Callaghan, J. M.; Booth, J. C. A large-signal model of ferroelectric thin-film transmission lines, *IEEE Trans. Microw. Theory Techn.*, **2011**, 12, pp. 3059-3067, doi: 10.1109/TMTT.2011.2169420.
71. Rezvanian, O.; et al, The role of creep in the time-dependent resistance of Ohmic gold contacts in radio frequency microelectromechanical system devices, *J. Appl. Phys.*, **2008**, 104, 024513, doi: 10.1063/1.2953072.
72. Zikry M. A. An accurate and stable algorithm for high strain-rate finite strain plasticity, Comput. Struct., 1994, 3 pp. 337-350.
73. Hyman D.; Mehregany M. Contact physics of gold microcontacts for MEMS switches, *IEEE Trans. Compon. Packag. Technol.*, **1999**, 22, pp. 357–364.
74. Holm R. Electric Contacts: Theory and Applications, **1967**, 4th edn., Berlin: Springer.